%% file: fmbird.tex
\documentclass{article}
\usepackage{setspace}

\usepackage{url}
\usepackage{natbib}
\usepackage{graphicx}
\usepackage{amsmath}

\providecommand{\methss}{\texttt{ss} }
\providecommand{\methrm}{\texttt{rm} }
\providecommand{\methmp}{\texttt{mp} }
\providecommand{\methdd}{\texttt{dd} }
\providecommand{\methssO}{\texttt{ss}}
\providecommand{\methrmO}{\texttt{rm}}
\providecommand{\methmpO}{\texttt{mp}}
\providecommand{\methddO}{\texttt{dd}}

\begin{document}


\title{Large-scale analysis of frequency modulation in birdsong databases}

\date{} 


\author{Dan Stowell and Mark D. Plumbley\\%
Centre for Digital Music, Queen Mary University of London\\
\texttt{dan.stowell@qmul.ac.uk}}

%

\maketitle 

\begin{abstract}
Birdsong often contains large amounts of rapid frequency modulation (FM). It is believed that the use or otherwise of FM is adaptive to the acoustic environment, and also that there are specific social uses of FM such as trills in aggressive territorial encounters. Yet temporal fine detail of FM is often absent or obscured in standard audio signal analysis methods such as Fourier analysis or linear prediction. %
Hence it is important to consider high resolution signal processing techniques for analysis of FM in bird vocalisations. %
If such methods can be applied at big data scales, this offers a further advantage as large datasets become available.

We introduce methods from the signal processing literature which can go beyond spectrogram representations to analyse the fine modulations present in a signal at very short timescales. Focusing primarily on the genus \textit{Phylloscopus}, we investigate which of a set of four analysis methods most strongly captures the species signal encoded in birdsong. In order to find tools useful in practical analysis of large databases, we also study the computational time taken by the methods, and their robustness to additive noise and MP3 compression. %

We find three methods which can robustly represent species-correlated FM attributes,
and that the simplest method tested also appears to perform the best.
We find that features representing the extremes of FM encode species identity supplementary to that captured in frequency features,
whereas bandwidth features do not encode additional information. %

Large-scale FM analysis can efficiently extract information useful for bioacoustic studies,
in addition to measures more commonly used to characterise vocalisations.
\end{abstract}



\section{Introduction}
\label{sec:intro}

Frequency modulation (FM) is an important component of much birdsong: various species of bird can discriminate the fine detail of frequency-chirped signals \citep{Dooling:2002,Lohr:2006}, and use fine FM information as part of their social interactions \citep{Trillo:2005,deKort:2009}.
Use of FM is also strongly species-dependent, in part due to adaptation of birds to their acoustic environment
\citep{Brumm:2009,Ey:2009}
.
Songbirds have specific musculature around the syrinx which endows them with independent fine control over frequency \citep{Goller:2012}.
They can control the two sides of their syrinx largely independently:
a sequence of two tones might be produced by each side separately, or by one side alone,
a difference shown by the absence/presence of brief FM ``slurs'' between notes \citep[e.g.\ Figure 9.8]{Marler:2004}.
Therefore, if we can analyse bird vocalisation recordings to characterise the use of FM across species and situations, this information could cast light upon acoustic adaptations and communicative issues in bird vocalisations.
As \citet{Slabbekoorn:2002} concluded, ``Measuring note slopes [FM], as well as other more traditional acoustic measures, may be important for comparative studies addressing these evolutionary processes in the future.''

Frequency analysis of birdsong is typically carried out using the short-time Fourier transform (STFT)
and displayed as a spectrogram.
FM can be observed implicitly in spectrograms, especially at slower modulation rates.
However, FM data are rarely explicitly quantified in bioacoustics analyses of birdsong (one exception is \citet{Gall:2012}), although the amount of FM is partly implicit in measurements such as the rate of syllables and the bandwidth (e.g. in \citet{Podos:1997,Vehrencamp:2013}).

The relative absence of fine FM analysis may be due to the difficulty in extracting good estimates of FM rates from spectrograms, especially with large data volumes.
Some previous work has indicated that the FM data extracted from a chirplet representation can improve the accuracy of a bird species classifier \citep{Stowell:2012c}.
However, there exists a variety of signal processing techniques which can characterise frequency-modulated sounds,
and no formal study has considered their relative merits for bird vocalisation analysis.

In the present work we aim to facilitate the use of direct FM measurements in bird bioacoustics, by conducting a formal comparison of four methods for characterising FM.
Each of these methods goes beyond the standard spectrogram analysis to capture detail of local modulations in a signal on a fine time-scale.
To explore the merits of these methods
we will use the machine learning technique of \textit{feature selection} \citep{Witten:2005} for a species classification task.

In the present work our focus is on methods that can be used with large bird vocalisation databases.
Many hypotheses about vocalisations could be explored using FM information, most fruitfully if data can be analysed at relatively large scale.
For this reason, we will describe an analysis workflow for audio which is simple enough to be fully automatic and to run across a large number of files. We will consider the runtime of the analysis techniques as well as the characteristics of the statistics they extract.

The genus \textit{Phylloscopus} (leaf warblers) has been studied previously for evidence of adaptive song variation.
For example \citet{Irwin:2008} studied divergence of vocalisation in a ring species (\textit{Phylloscopus trochiloides}),
suggesting that stochastic genetic drift may be a major factor in diversity of vocalisations.
\citet{Mahler:2009} found correlations between aspects of frequency range and body size across the \textit{Phylloscopus} genus.
They also considered character displacement effects, which one might expect to cause the song of sympatric species to diverge,
but found no significant such effect on the song features they measured.
\citet{Linhart:2012b} studied \textit{Phylloscopus collybita},
also finding a connection between song frequency and body size.
Such research context motivated our choice to use \textit{Phylloscopus} as
our primary focus in this study, in order to develop signal analysis methods that might provide further data on song structure.
However, we also conducted a larger-scale FM analysis using a database with samples representing species across the wider order of Passeriformes.
We first discuss the FM analysis methods to be considered.

\subsection{FM analysis methods}
\label{sec:fmmethods}

For many purposes, the standard representation of audio signals is the spectrogram,
calculated from the magnitudes of the windowed short-time Fourier transform (STFT).
The STFT is applied to each windowed ``frame'' of the signal (of duration typically 10 or 20 ms),
resulting in a representation of variations across time and frequency.
The spectrogram is widely used in bioacoustics, and a wide variety of measures are derived from this, manually or automatically:
it is common to measure the minimum and maximum frequencies in each recording or each syllable, as well as durations, amplitudes and so forth \citep{Marler:2004}.
Notable for the present work is the FM rate measure of \citet{Gall:2012}, derived from manual identification of frequency inflection points (i.e.\ points at which the modulation changes from upward to downward, or downward to upward) on a spectrogram.
\citet{Trillo:2005} characterise ``trill vigour'' in a related manner but applicable only to trilled syllables.
For fully automatic analysis, in Section \ref{sec:method} we will describe a method related to that of \citet{Gall:2012} but with no manual intervention.

The spectrogram is a widespread tool, but it does come with some limitations.
Analysing a 10 or 20 ms frame with the STFT
implies the assumption that the signal is \textit{locally stationary}
(or \textit{pseudo-stationary}),
meaning it is produced by a process whose parameters (such as the fundamental frequency)
do not change across the duration of the individual frame \citep[Section 10.6.3]{Mallat:1999}.
However, many songbirds sing with very dramatic and fast FM (as well as AM),
which may mean that the local stationarity assumption is violated and that there is fine-resolution FM which cannot be represented with a standard spectrogram.

Signal analysis is under-determined in general:
many different processes can in principle produce the same audio signal.
Hence the representations derived by STFT and LPC analysis are but two families of possible ``explanation'' for the observed signal.
A large body of research in signal processing has considered alternative representations, tailored to various classes of signal including signals with fast FM.
One recent example which was specifically described in the context of birdsong is that of \citet{Stowell:2012c},
which uses a kind of \textit{chirplet analysis} to add an extra chirp-rate dimension to a spectrogram.
A ``chirplet'' is a short-time packet of signal having a central frequency, amplitude, and a parametric chirp-rate which modulates the frequency over time.
More generally, the field of \textit{sparse representations} allows one to define a ``dictionary'' of a large number of elements from which a signal may be composed, and then to analyse the signal into a small number of components selected from the dictionary \citep{Plumbley:2010}.
For the present purposes, notable is the method of \citet{Gribonval:2001} which applies an accelerated version of a technique known as \textit{Matching Pursuit} specifically adapted to analyse a signal as a sparse combination of chirplets.

Alternative paradigms are also candidates for performing high-resolution FM analysis.
One paradigm is that of \textit{spectral reassignment},
based on the idea that after performing an STFT analysis
it is possible to ``reassign'' the resulting list of frequencies and magnitudes
to shift them to positions which are in some sense a better fit to the evidence
\citep{Fulop:2006}.
The \textit{distribution derivative} method (DDM) of \citet[Chapter 10]{Musevic:2013}
is one such approach which is able to reassign a spectrum to find the best-matching parameters on the assumption that the signal is composed of
amplitude- and frequency-modulated sinusoids.

Another approach is that of \citet{Badeau:2006} which uses a subspace model to achieve high-resolution characterisation of signals with smooth modulations.
However, there may be limitations on the rate of FM that can be reflected faithfully:
this method relies on a smoothness assumption in the frame-to-frame evolution of the sound which means that it is most suited to relatively moderate rates of FM, such as the vibrato in human singing.

In the following we will apply a selection of analysis techniques to birdsong recordings,
and study whether the FM information extracted is a reliable signal of species identity.
This is not the only application for which FM information is relevant:
our aim is that this exploration will encourage other researchers to add high-resolution FM analysis to their toolbox.

\section{Materials and methods}
\label{sec:expt}
\subsection{Data}
\label{sec:data}

We first collected a set of recordings of birds in the genus \textit{Phylloscopus} from a dataset made available by the Animal Sound Archive in Berlin.%
\footnote{\url{http://www.animalsoundarchive.org/}}
This consisted of 45 recordings over 5 species, in WAV format, with durations ranging from 34 seconds to 19 minutes.
In the following we will refer to this dataset as \textbf{\textit{PhyllASA}}.

As a second dataset, we also considered a broader set of audio from the Animal Sound Archive, not confined to \textit{Phylloscopus} but across the order \textit{Passeriformes} (762 recordings over 84 species).
We will refer to this as \textbf{\textit{PassaASA}}.

Thirdly we collected a larger \textit{Phylloscopus} dataset from the online archive Xeno Canto.%
\footnote{\url{http://www.xeno-canto.org/}}
This consisted of 1390 recordings across 56 species, ranging widely in duration from one second to seven minutes.
Our criteria for selecting files from the larger Xeno Canto archive were:
genus \textit{Phylloscopus};
quality level A or B (the top two quality ratings);
not flagged as having uncertain species identity.
In the following we will refer to this dataset as \textbf{\textit{PhyllXC}}.

Note that the ``crowdsourced'' Xeno Canto dataset is qualitatively different from \textit{PhyllASA}.
Firstly it was compiled from various contributors online, and so is not as tightly controlled.
The noise conditions and recording quality can vary widely.
Secondly, all audio content is compressed in MP3 format (with original uncompressed audio typically unavailable).
The MP3 format reduces filesize by discarding information which is considered unnecessary for audio quality as judged by human perception \citep{mp3:1993}.
However, human and avian audition differ in important ways, including time and frequency resolution,
and we cannot assume that MP3 compression is ``transparent'' regarding the species-specific information that might be important in bird communication.
Hence in our study we used this large crowdsourced MP3 dataset
only after testing experimentally the impact of compression and signal degradation on the features we measured (using the \textit{PhyllASA} data).

For each dataset considered here, we resampled audio files to 48 kHz mono WAV format before processing, and truncated long files to a maximum duration of 5 minutes.
All of the datasets contain an uneven distribution, with some species represented in more recordings than others (Table \ref{tbl:specieslists}).
This is quite common but carries implications for the evaluation of automatic classification, as will be discussed below.

\input{specieslists}

\subsection{Method}
\label{sec:method}

For all analysis methods we used a frame size of 512 samples (10.7 milli\-seconds, at 48 kHz), with Hann windowing for STFT,
and the frequency range of interest was restricted to 2--10 kHz.
For each recording in each dataset, we applied a fully automatic analysis using each of four signal processing techniques.
Our requirement of full automation excludes a preprocessing step of manually segmenting of birdsong syllables from the background.
We chose to use the simplest form of automatic segmentation, simply to select the 10\% of highest-energy frames in each recording.
More sophisticated procedures can be applied in future; however, as well as simplicity this method has an advantage of speed when analysing large databases.
We analysed each recording using each of the following techniques (which we assign two-letter identifiers for reference):

\begin{description}
    \item[\methssO:]
	a spectrographic method
 related to the method of \citet{Gall:2012} but with no manual intervention, as follows.
Given a sample of birdsong, for every temporal frame we identify the frequency having peak energy, within the frequency region of interest.
We calculate the absolute value of the first difference, i.e.\ the magnitude of the frequency jump between successive frames.
We then summarise this by the median or other statistics, to characterise the distribution over the depth of FM present in each recording.
This method relies on the peak-energy within each frame rather than manual identification of inflection points in the pitch trace,
which means that it is potentially susceptible to noise and other corruptions,
but it remains a relatively robust technique which can be applied to a standard spectrogram representation.
In the following we will refer to this method as the ``simple spectrographic'' method.
    \item[\methrmO:]
	the heterodyne (ring modulation) chirplet analysis of \citet{Stowell:2012c}, taking information from the peak-energy detection in each frame.%
\footnote{Python source code for the method of \citet{Stowell:2012c} is available at \newline \url{https://code.soundsoftware.ac.uk/projects/chirpletringmod}.}
    \item[\methmpO:]
	the Matching Pursuit technique of \citet{Gribonval:2001}, implemented using the open-source Matching Pursuit ToolKit (MPTK) v0.7.%
		\footnote{Available at \url{http://mptk.irisa.fr/}.}
	For this technique the 10\% highest-energy threshold is not applicable, since the method is iterative and could return many more results than there are signal frames: we automatically set a threshold at a number of results which recovers roughly the same amount of signal as the 10\% threshold.
    \item[\methddO:]
        the distribution derivative method (DDM) of
        \citet[Chapter 10]{Musevic:2013}, taking information from the peak-energy sinusoid detected in each frame.%
\footnote{Matlab/Octave source code for the method of \citet{Musevic:2013} is available at \newline \url{https://code.soundsoftware.ac.uk/projects/ddm}.}
\end{description}
We also conducted a preliminary test with the subspace method of \citet{Badeau:2006}, but this proved to be inappropriate for the rapid FM modulations found in birdsong because of an assumption of smooth FM variation inherent in the method (Badeau, pers. comm.).

Each of these methods resulted in a list of ``frames'' or ``atoms'' for a recording, each with an associated frequency and FM rate.
In order to characterise each recording as a whole,
we selected summary statistics over these frames in a recording to use as features.
We summarised the frequency data by their median, and by their 5- and 95-percentiles.
The 5- and 95-percentiles are robust measures of minimum and maximum frequency;
we also calculated the ``bandwidth'' as the difference between the 5- and 95-percentile.
We summarised the FM data by their median, and also by their 75- and 95-percentiles.
These percentiles were chosen to explore whether information about the relative extremes of FM found in the recording provide useful extra information.

So, for each recording and each analysis method we can extract a set of frequency and FM summary features.
It remains to determine which of these features might be most useful in looking for signals of species identity in recorded bird vocalisations.
We explored this through two interrelated approaches: feature selection, and automatic classification experiments.
Through these two approaches, we were able to compare the different features against each other,
and also compare the features as extracted by each of the four signal-processing techniques given above.

One approach that has been used to explore the value of different features is \textit{principal components analysis} (PCA) applied to the features,
to determine axes that represent the strongest dimensions of variance in the features
(see e.g.\ \citet{Mahler:2009,Handford:1991}).
This method is widespread and well-understood.
However, it is a purely linear analysis which may fail to reflect nonlinear information-carrying patterns in the data;
and more importantly for our purposes, PCA does not take into account the known species labels,
and so can only ever serve as indirect illumination on questions about which features might carry such information.

In the field of data mining/machine learning, researchers instead use \textit{feature selection} techniques to evaluate directly
the predictive power that a feature (or a set of features) has with respect to some attribute \citep{Witten:2005}.
We used an information-theoretic feature selection technique from that field.
In \textit{information gain} feature selection, each of our features is evaluated by measuring the information gain with respect to the species label, which is the amount by which the feature reduces our uncertainty in the label:

\begin{equation*}
	\text{IG}(\text{Species}, \text{Feature}) = H(\text{Species}) - H(\text{Species} | \text{Feature})
\end{equation*}
where $H(\cdot)$ is the Shannon entropy.
The value $H(\text{Species})$ represents the number of binary bits of information that must typically be conveyed in order to identify the species of an individual (from a fixed set of species).
The information gain $\text{IG}(\text{Species}, \text{Feature})$ then tells us how many of those binary bits are already encoded in a particular feature, i.e.\ the extent to which that feature reduces the uncertainty of the species identity.
If a feature is repeatedly ranked highly, this means that it contains a stronger signal of species identity than lower-ranked features and thus suggests it should be a useful measure.
The approach just described is reminiscent of the information-theoretic method introduced by \citet{Beecher:1989},
except that his concern was with signals of individual identity rather than species identity.

Having performed feature selection, we were then able to choose promising subsets of features which might concisely represent species information.
To evaluate these subsets concretely we conducted
an experiment in automatic species classification.
For this we
used a leading classification algorithm, the Support Vector Machine (SVM),
implemented in the \textit{libsvm} library version 3.1, choosing the standard radial basis function SVM classifier.
The evaluation statistic we used was the weighted ``area under the receiver operating characteristics curve'' (the weighted \textit{AUC}),
which summarises the rates of true-positive and false-positive detections made
\citep{Fawcett:2006}.
This measure is more appropriate than raw accuracy, when analysing datasets with wide variation in numbers per class as in the present case (ibid.).
The AUC yields the same information as the Wilcoxon signed-rank statistic \citep{Hanley:1982}.
The feature selection and classification experiments were all performed using \textit{Weka} 3.6.0 \citep{Witten:2005},
and analysed using \textit{R} version 2.13.1 \citep{R:2010}.

An important issue when considering automatic feature extraction is the robustness of the features to corruptions that may be found in audio databases, such as background noise or MP3 compression artifacts.
This has particular pertinence for the crowdsourced \textit{PhyllXC} dataset, as discussed above.
For this reason, we also studied our first dataset after putting the audio files through two corruption processes: added white noise ($-45$ dB relative to full-scale, judged by ear to be noticeable but not overwhelming), and MP3 compression (64 kbps, using the \textit{lame} software library version 3.99.5).
To quantify whether an audio feature was badly impacted by such corruption, we measured the Pearson correlations of the features measured on the original dataset with their corrupted equivalent.
This test does not depend on species identity as in our main experimental tests, but simply on the numerical stability of the summary statistics we consider.

In this study we focussed on frequency and FM characteristics of sounds, both of which can be extracted completely automatically from short time frames.
We did not include macro-level features such as syllable lengths or syllable rates,
because reliable automatic extraction of these is complex.
Rather, we compared the fine-detail FM analyses against frequency measures, the latter being common in the bioacoustics literature:
our feature set included features corresponding to the lower, central and upper frequency, and frequency bandwidth.

\input{results}

\section{Conclusions}
\label{sec:conc}

In much research involving acoustic analysis of birdsong,
frequency modulation (FM) has been measured manually,
described qualitatively or left implicit in other measurements such as bandwidth.
We have demonstrated that it is possible to extract
data about FM on a fine temporal scale,
from large audio databases, in fully automatic fashion,
and that this data encodes aspects of ecologically pertinent information such as species identity.
Further, we have demonstrated that a relatively simple technique based on spectrogram data
is sufficient to extract information pertinent to species,
which one might expect could only be extracted with more advanced signal-processing techniques.
Our study provides evidence that researchers can and should measure such FM characteristics
when analysing the acoustic characteristics of bird vocalisations.

\section*{Acknowledgments}
%
DS \& MP are supported by an EPSRC Leadership Fellowship EP/G007144/1.
Our thanks to: Alan McElligott for helpful advice while preparing the manuscript;
Sa\v{s}o Mu\v{s}evi\v{c} for discussion and for making his DDM software available;
and R\'{e}mi Gribonval and team at INRIA Rennes for discussion
and software development during a research visit.

\section*{Data accessibility}

The feature values for each sound file are available in online data tables.%
\footnote{\url{http://dx.doi.org/10.6084/m9.figshare.795273}}
The original audio for the \textit{PhyllXC} dataset can be retrieved from the Xeno Canto website,
using the XC ID numbers given in the online data table.
The original audio for the \textit{PhyllASA} and \textit{PassaASA} datasets can be requested from the Animal Sound Archive,
using the track filenames given in the online data table.

\bibliographystyle{bes}
\bibliography{../refs}

\input{extraresults}

\end{document}

%% file: specieslists.tex
\begin{table}[p]
\caption{Counts of species occurrence in our three datasets. Note that \textit{PhyllASA} is a subset of \textit{PassaASA}, as reflected in the counts.%
}
\label{tbl:specieslists}

\resizebox{!}{0.35\textheight}{
\begin{tabular}{l | r r r}
Species	&	\rotatebox{90}{PhyllASA}	&	\rotatebox{90}{PassaASA}	&	\rotatebox{90}{PhyllXC} \\
\hline
Acrocephalus arundinaceus	&	 	&	9	&	 	\\
Acrocephalus palustris	&	 	&	12	&	 	\\
Acrocephalus schoenobaenus	&	 	&	3	&	 	\\
Acrocephalus scirpaceus	&	 	&	5	&	 	\\
Aegithalos caudatus	&	 	&	1	&	 	\\
Alauda arvensis	&	 	&	8	&	 	\\
Anthus pratensis	&	 	&	1	&	 	\\
Anthus trivialis	&	 	&	74	&	 	\\
Apalis chariessa	&	 	&	3	&	 	\\
Calcarius lapponicus	&	 	&	1	&	 	\\
Carduelis carduelis	&	 	&	1	&	 	\\
Carduelis chloris	&	 	&	3	&	 	\\
Carduelis spinus	&	 	&	4	&	 	\\
Certhia brachydactyla	&	 	&	3	&	 	\\
Certhia familiaris	&	 	&	1	&	 	\\
Corvus corax	&	 	&	1	&	 	\\
Corvus corone	&	 	&	3	&	 	\\
Cyanocitta cristata	&	 	&	2	&	 	\\
Delichon urbica	&	 	&	4	&	 	\\
Emberiza calandra	&	 	&	4	&	 	\\
Emberiza citrinella	&	 	&	34	&	 	\\
Emberiza hortulana	&	 	&	94	&	 	\\
Emberiza pusilla	&	 	&	1	&	 	\\
Emberiza rustica	&	 	&	3	&	 	\\
Emberiza schoeniclus	&	 	&	11	&	 	\\
Emberiza spodocephala	&	 	&	2	&	 	\\
Erithacus rubecula	&	 	&	14	&	 	\\
Ficedula albicollis	&	 	&	1	&	 	\\
Ficedula hypoleuca	&	 	&	4	&	 	\\
Ficedula parva	&	 	&	7	&	 	\\
Fringilla coelebs	&	 	&	87	&	 	\\
Fringilla montifringilla	&	 	&	9	&	 	\\
Garrulax subunicolor	&	 	&	1	&	 	\\
Garrulus glandarius	&	 	&	2	&	 	\\
Hippolais icterina	&	 	&	19	&	 	\\
Hirundo rustica	&	 	&	3	&	 	\\
Lanius collurio	&	 	&	4	&	 	\\
Locustella fluviatilis	&	 	&	5	&	 	\\
Locustella lanceolata	&	 	&	1	&	 	\\
Locustella luscinioides	&	 	&	3	&	 	\\
Locustella naevia	&	 	&	6	&	 	\\
Loxia curvirostra	&	 	&	1	&	 	\\
Lullula arborea	&	 	&	6	&	 	\\
Luscinia calliope	&	 	&	 	&	 	\\
Luscinia luscinia	&	 	&	10	&	 	\\
Luscinia megarhynchos	&	 	&	26	&	 	\\
Luscinia svecica	&	 	&	3	&	 	\\
\end{tabular}

\begin{tabular}{l | r r r}
Species	&	\rotatebox{90}{PhyllASA}	&	\rotatebox{90}{PassaASA}	&	\rotatebox{90}{PhyllXC} \\
\hline
Motacilla alba	&	 	&	1	&	 	\\
Motacilla flava	&	 	&	3	&	 	\\
Muscicapa striata	&	 	&	1	&	 	\\
Nucifraga caryocatactes	&	 	&	20	&	 	\\
Panurus biarmicus	&	 	&	1	&	 	\\
Parus ater	&	 	&	5	&	 	\\
Parus caeruleus	&	 	&	8	&	 	\\
Parus major	&	 	&	9	&	 	\\
Parus montanus	&	 	&	4	&	 	\\
Parus palustris	&	 	&	3	&	 	\\
Perisoreus infaustus	&	 	&	1	&	 	\\
Phoenicurus ochruros	&	 	&	3	&	 	\\
Phoenicurus phoenicurus	&	 	&	22	&	 	\\
Phylloscopus affinis	&	 	&	 	&	7	\\
Phylloscopus amoenus	&	 	&	 	&	2	\\
Phylloscopus armandii	&	 	&	 	&	6	\\
Phylloscopus bonelli	&	3	&	3	&	71	\\
Phylloscopus borealis	&	 	&	 	&	25	\\
Phylloscopus borealoides	&	 	&	 	&	1	\\
Phylloscopus budongoensis	&	 	&	 	&	1	\\
Phylloscopus calciatilis	&	 	&	 	&	9	\\
Phylloscopus canariensis	&	 	&	 	&	11	\\
Phylloscopus cantator	&	 	&	 	&	6	\\
Phylloscopus cebuensis	&	 	&	 	&	4	\\
Phylloscopus chloronotus	&	 	&	 	&	10	\\
Phylloscopus claudiae	&	 	&	 	&	15	\\
Phylloscopus collybita	&	12	&	12	&	323	\\
Phylloscopus coronatus	&	 	&	 	&	6	\\
Phylloscopus davisoni	&	 	&	 	&	2	\\
Phylloscopus emeiensis	&	 	&	 	&	4	\\
Phylloscopus examinandus	&	 	&	 	&	3	\\
Phylloscopus forresti	&	 	&	 	&	14	\\
Phylloscopus fuligiventer	&	 	&	 	&	4	\\
Phylloscopus fuscatus	&	5	&	5	&	33	\\
Phylloscopus griseolus	&	 	&	 	&	6	\\
Phylloscopus hainanus	&	 	&	 	&	3	\\
Phylloscopus herberti	&	 	&	 	&	4	\\
Phylloscopus humei	&	 	&	 	&	51	\\
Phylloscopus ibericus	&	 	&	 	&	42	\\
Phylloscopus ijimae	&	 	&	 	&	2	\\
Phylloscopus inornatus	&	 	&	 	&	53	\\
Phylloscopus kansuensis	&	 	&	 	&	4	\\
Phylloscopus laetus	&	 	&	 	&	1	\\
Phylloscopus maculipennis	&	 	&	 	&	16	\\
Phylloscopus magnirostris	&	 	&	 	&	13	\\
Phylloscopus makirensis	&	 	&	 	&	7	\\
Phylloscopus neglectus	&	 	&	 	&	3	\\
\end{tabular}

\begin{tabular}{l | r r r}
Species	&	\rotatebox{90}{PhyllASA}	&	\rotatebox{90}{PassaASA}	&	\rotatebox{90}{PhyllXC} \\
\hline
Phylloscopus nigrorum	&	 	&	 	&	7	\\
Phylloscopus nitidus	&	 	&	 	&	9	\\
Phylloscopus occisinensis	&	 	&	 	&	5	\\
Phylloscopus ogilviegranti	&	 	&	 	&	15	\\
Phylloscopus olivaceus	&	 	&	 	&	2	\\
Phylloscopus orientalis	&	 	&	 	&	5	\\
Phylloscopus plumbeitarsus	&	 	&	 	&	10	\\
Phylloscopus poliocephalus	&	 	&	 	&	8	\\
Phylloscopus presbytes	&	 	&	 	&	15	\\
Phylloscopus proregulus	&	 	&	 	&	17	\\
Phylloscopus pulcher	&	 	&	 	&	6	\\
Phylloscopus reguloides	&	 	&	 	&	26	\\
Phylloscopus ricketti	&	 	&	 	&	1	\\
Phylloscopus ruficapilla	&	 	&	 	&	7	\\
Phylloscopus sarasinorum	&	 	&	 	&	11	\\
Phylloscopus schwarzi	&	 	&	 	&	16	\\
Phylloscopus sibilatrix	&	11	&	11	&	105	\\
Phylloscopus sindianus	&	 	&	 	&	7	\\
Phylloscopus subviridis	&	 	&	 	&	1	\\
Phylloscopus tenellipes	&	 	&	 	&	28	\\
Phylloscopus trivirgatus	&	 	&	 	&	16	\\
Phylloscopus trochiloides	&	 	&	 	&	61	\\
Phylloscopus trochilus	&	14	&	14	&	208	\\
Phylloscopus tytleri	&	 	&	 	&	1	\\
Phylloscopus umbrovirens	&	 	&	 	&	5	\\
Phylloscopus xanthoschistos	&	 	&	 	&	25	\\
Phylloscopus yunnanensis	&	 	&	 	&	11	\\
Prunella modularis	&	 	&	2	&	 	\\
Pyrrhula pyrrhula	&	 	&	1	&	 	\\
Regulus ignicapillus	&	 	&	3	&	 	\\
Regulus regulus	&	 	&	2	&	 	\\
Saxicola rubetra	&	 	&	2	&	 	\\
Sitta europaea	&	 	&	6	&	 	\\
Smithornis capensis	&	 	&	1	&	 	\\
Sturnus vulgaris	&	 	&	1	&	 	\\
Sylvia atricapilla	&	 	&	14	&	 	\\
Sylvia borin	&	 	&	10	&	 	\\
Sylvia communis	&	 	&	9	&	 	\\
Sylvia curruca	&	 	&	2	&	 	\\
Sylvia nisoria	&	 	&	2	&	 	\\
Troglodytes troglodytes	&	 	&	11	&	 	\\
Turdus iliacus	&	 	&	2	&	 	\\
Turdus merula	&	 	&	36	&	 	\\
Turdus philomelos	&	 	&	21	&	 	\\
Turdus pilaris	&	 	&	4	&	 	\\
Turdus viscivorus	&	 	&	7	&	 	\\
	&	 	&		&	 	\\
\end{tabular}
}
\end{table}

%% file: results.tex
\section{Results}
\label{sec:results}

\begin{figure}[t]
	\centering
	\includegraphics [width=0.89\textwidth,clip,trim=15mm 5mm 20mm 9mm]  {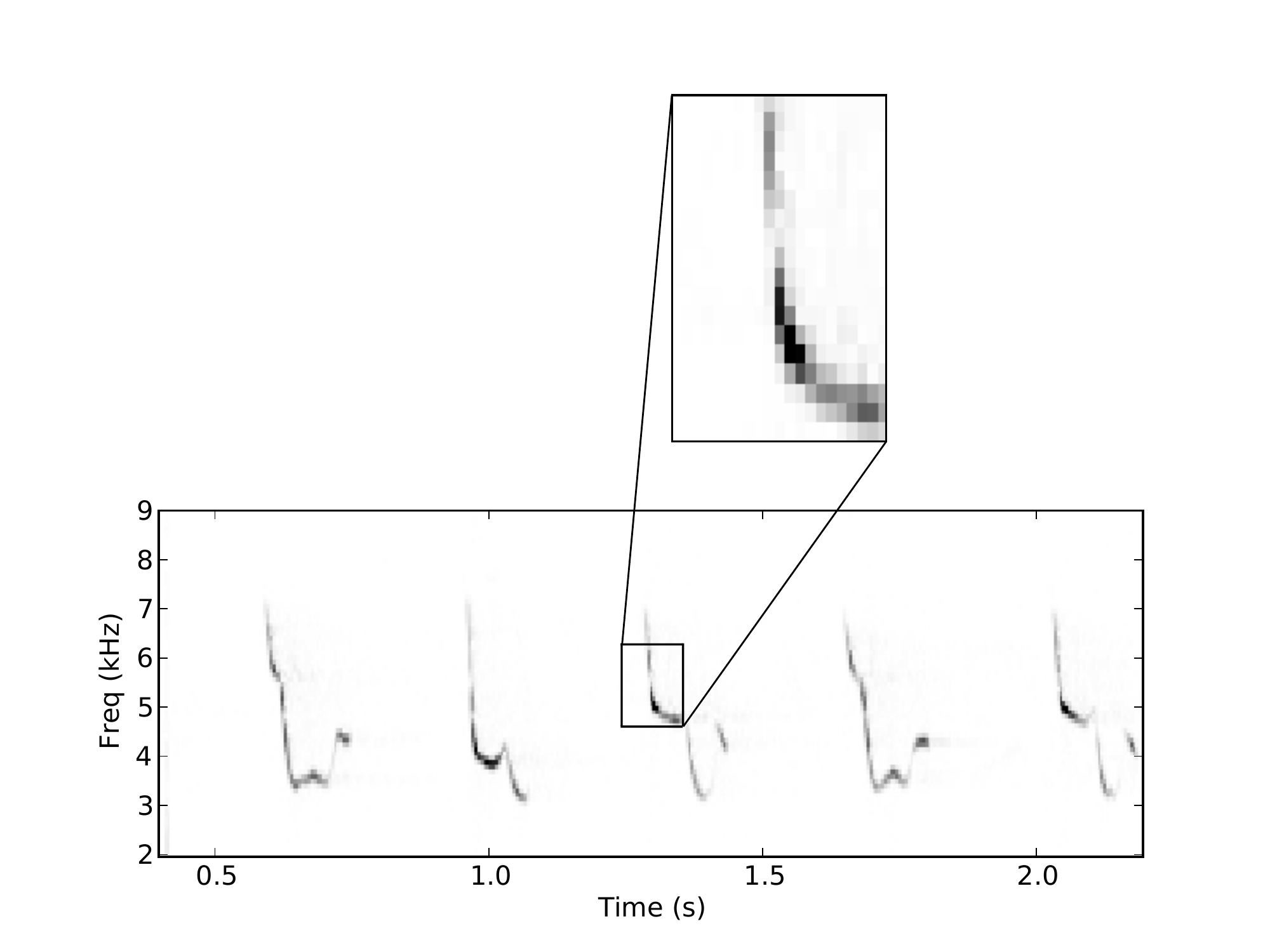}
	\caption{Standard spectrogram for a short excerpt of Chiffchaff (\textit{Phylloscopus collybita}). %
	The FM can be seen by eye but is not explicit in the underlying data, being spread across many ``pixels''. %
	}
\label{fig:specgram}
\end{figure}

\begin{figure}[pt]
	\centering
	\includegraphics [width=0.89\textwidth,clip,trim=15mm 81mm 20mm 9mm]  {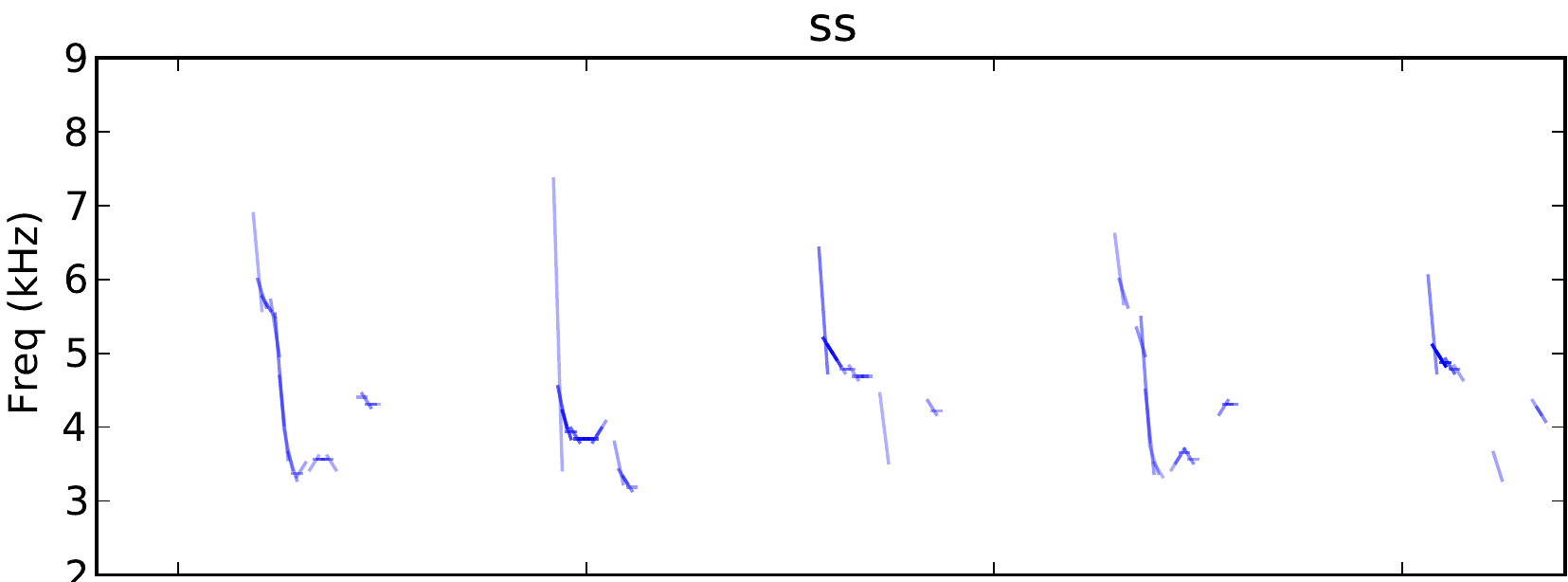}
	\includegraphics [width=0.89\textwidth,clip,trim=15mm 81mm 20mm 9mm]  {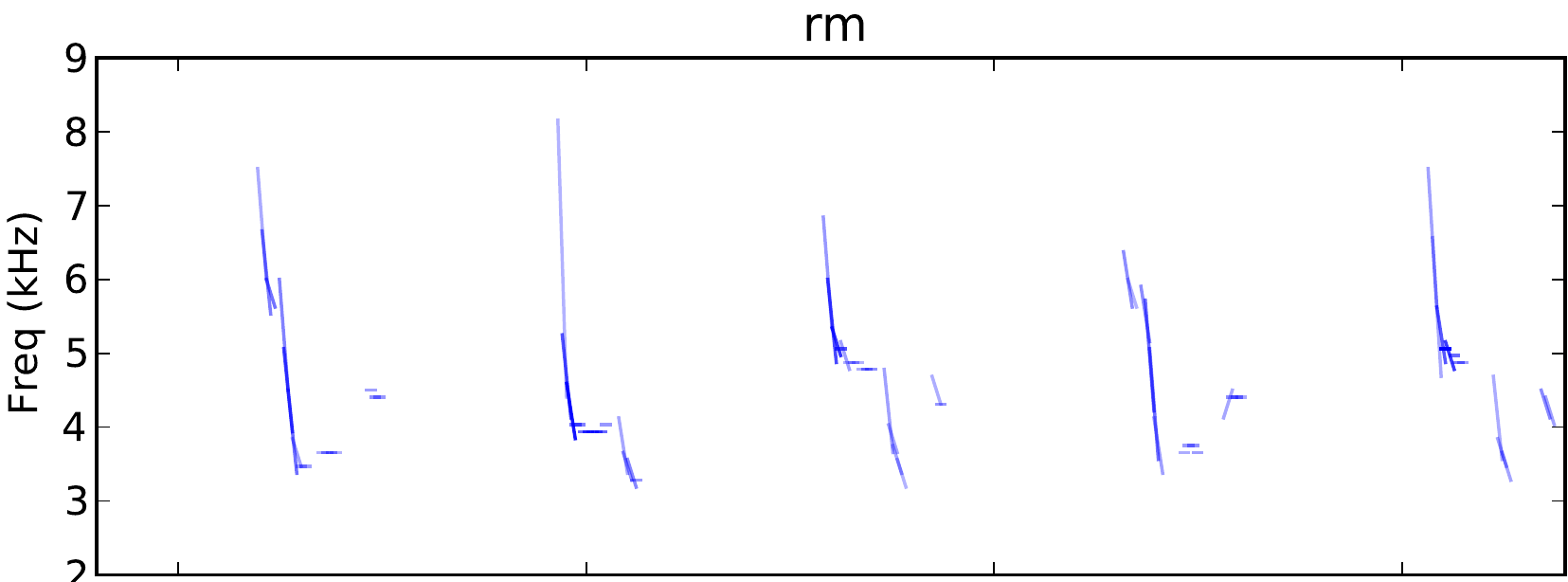}
	\includegraphics [width=0.89\textwidth,clip,trim=15mm 81mm 20mm 9mm]  {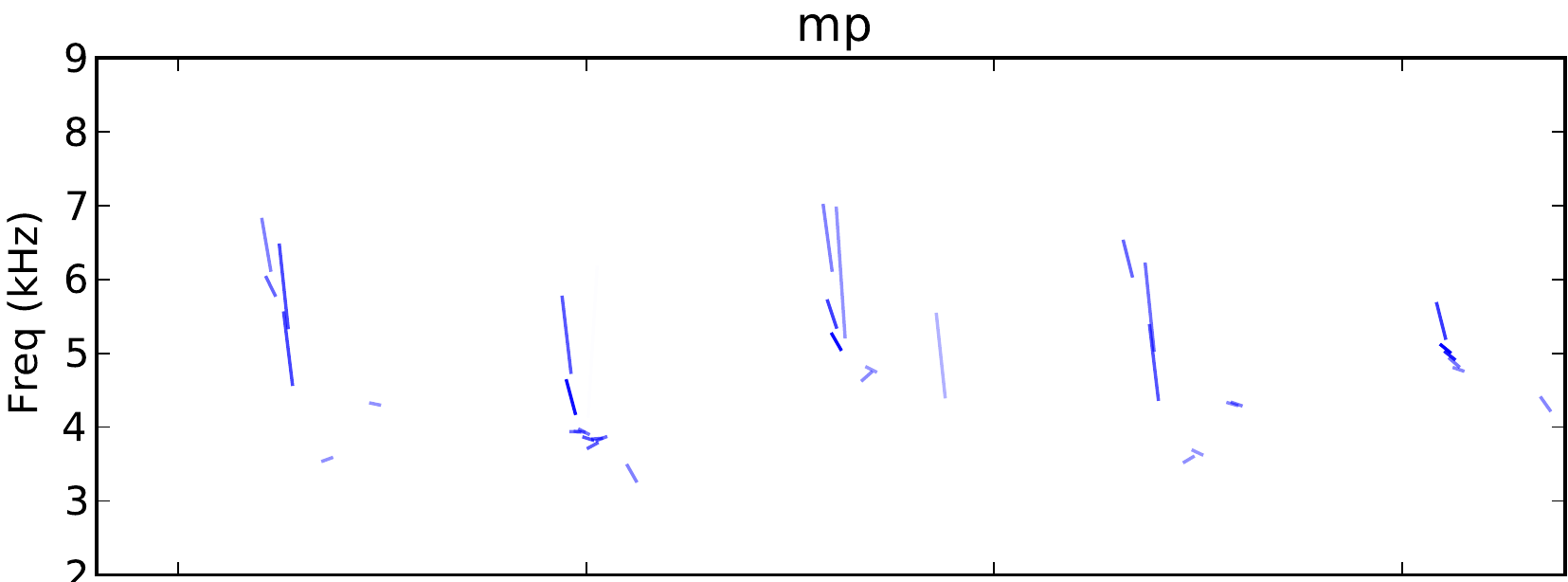}
	\includegraphics [width=0.89\textwidth,clip,trim=15mm 70mm 20mm 9mm]  {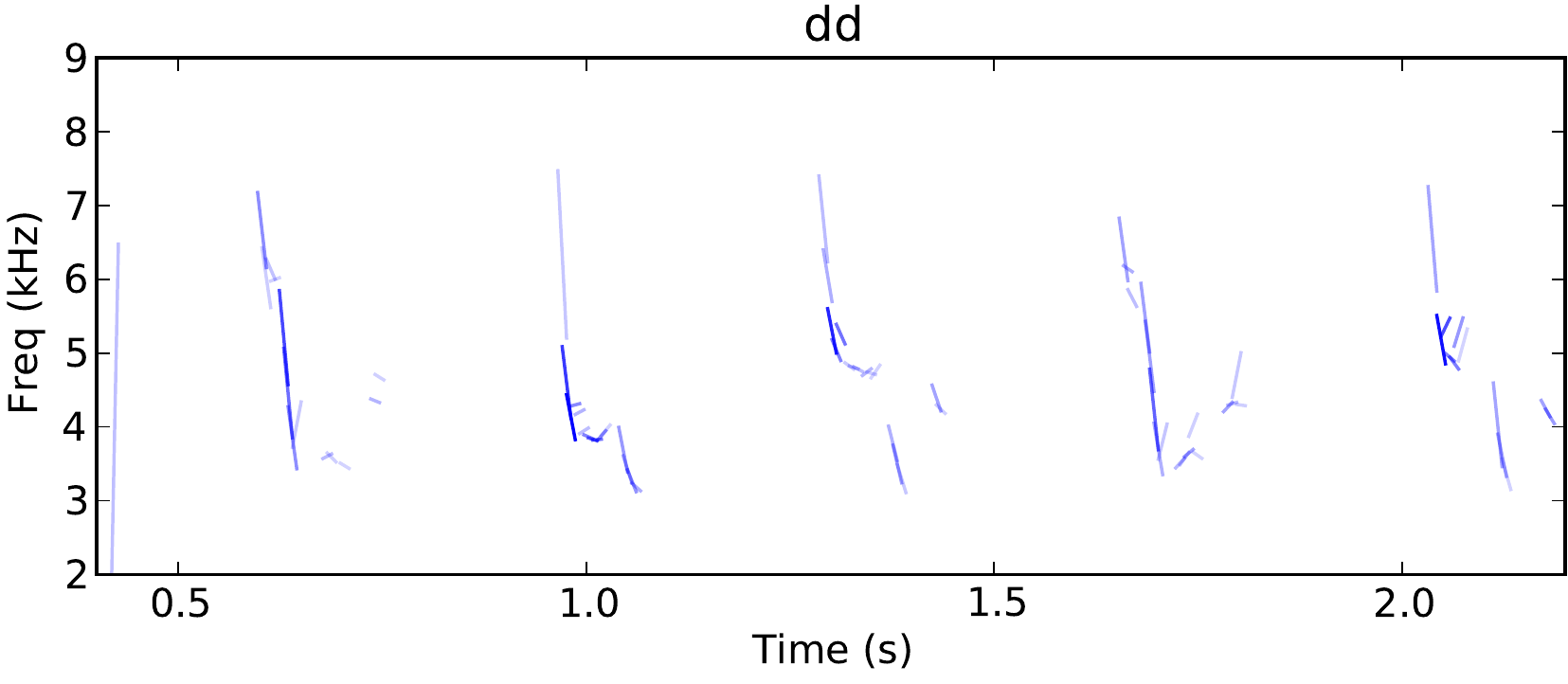}
	\caption{Time-frequency plots of the ``chirp'' data recovered by each method, for the same excerpt as in Figure \ref{fig:specgram}.
	}
\label{fig:chirpograms}
\end{figure}

We first illustrate the data which is produced by the analysis methods tested,
using a recording of \textit{Phylloscopus collybita} (Chiffchaff) from \textit{PhyllASA} as an example.
Figure \ref{fig:specgram} shows a conventional spectrogram plot for our chosen excerpt.
We can infer FM characteristics visually, but the underlying data (a grid of intensity ``pixels'') does not directly present FM for analysis.
Figure \ref{fig:chirpograms} represents the same excerpt analysed by each of the methods we consider.
Each of the plots appears similar to a conventional spectrogram,
showing the presence of energy at particular time and frequency locations.
However, instead of a uniform grid
the image is created from a set of line segments,
each segment having a location in time and frequency but also a slope.
It is clear from Figure \ref{fig:chirpograms} that each of the methods can build up a portrait of the birdsong syllables,
although some are more readable than others.
The plot from \methmp appears more fragmented than the others.
This can be traced back to the details of the method used,
but for now we merely note that the apparent neatness of each representation does not necessarily indicate which method
most usefully captures species-specific FM characteristics.

 \begin{table}[t]
\caption{Time taken to run each analysis method on our first dataset \mbox{\textit{PhyllASA}}, expressed as a proportion of the total duration of the audio files %
(so that any number below 1 indicates faster than real-time processing). %
Times were measured on a laptop with Intel i5 2.5 GHz processor. %
}
\label{tbl:timings}
	\centering
\begin{tabular}{l l}
	Method & Time taken (relative to audio duration) \\
\hline
ss	&	0.02 \\
rm	&	0.40 \\
mp	&	0.58 \\
dd	&	1.22 \\
\hline
\end{tabular}
\end{table}

The relative speeds of the analysis methods described here are given in Table \ref{tbl:timings}.
The simple spectrogram method is by far the fastest,
as is to be expected given its simplicity.
All but one of the methods run much faster than realtime,
though the difference in speed between the simple spectrogram and the more advanced methods is notable,
and certainly pertinent when considering the analysis of large databases.

\begin{figure}[t]
	\centering
	\includegraphics [width=0.7\textwidth,clip,trim=10mm 5mm 20mm 13mm]  {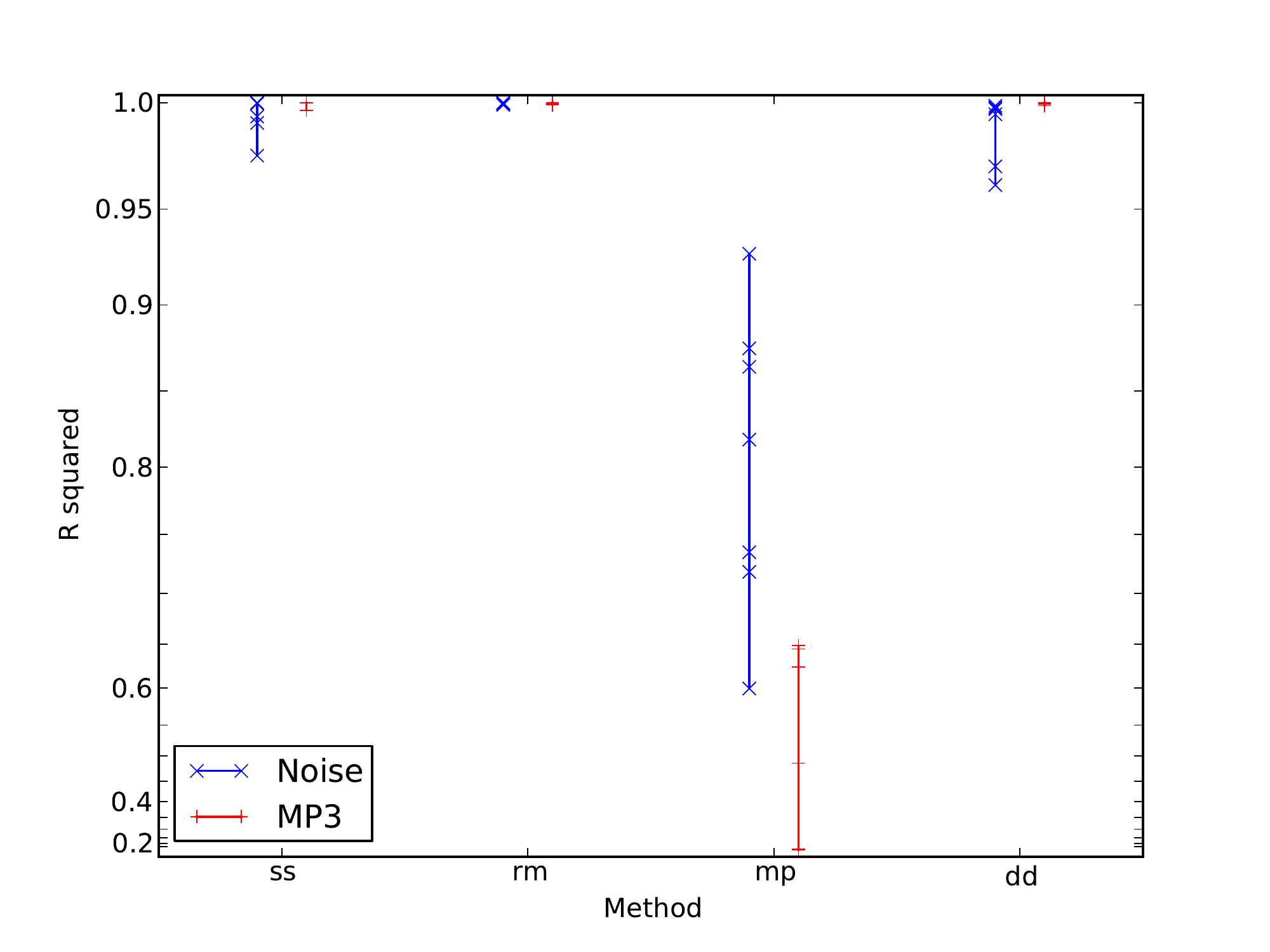}
\caption{Squared Pearson correlation between audio features and their values after applying audio degradation, across the \textit{PhyllASA} dataset. %
Each point represents one feature; features are grouped by analysis method and degradation type. %
We inspected the variation according to feature, and found no general tendencies; %
therefore features are collapsed into a single column per analysis method in order to visualise the differences in range. %
Note that the vertical axis is warped to enhance visibility at the top end of the scale. %
}
\label{fig:corr}
\end{figure}

Features extracted by methods \methss \methrm and \methdd were highly robust to the noise and MP3 degradations applied,
in all cases having a correlation with the original features better than 0.95 (Figure \ref{fig:corr}).
Method \methrm showed particularly strong robustness.
The \methmp method, on the other hand,
yielded features of very low robustness:
correlation with the original features was never above 0.95, in some cases going as low as to be around zero.
This indicates that features from the \methmp method may be generally unreliable when applied to the \textit{PhyllXC} dataset considered next.

Our feature selection experiments revealed notable trends in the information gain (IG) values associated with certain features,
with broad commonalities across the three datasets tested
(see Appendix for details).
In particular, the bandwidth features achieve very low IG values in all cases.
Conversely, the median frequency feature performs strongly for all datasets and all methods.
The FM features perform relatively strongly on \textit{PhyllASA}, appearing generally stronger than frequency features,
but this pattern does not persist into the other (larger) datasets.
However, the 75-percentile of FM did generally rank highly in the feature selection results.

Based on the results of feature selection, we chose to take the following four feature sets forward to the classification experiment:
\begin{itemize}
	\item	Three FM features (\texttt{fm\_med}, \texttt{fm\_75pc}, \texttt{fm\_95pc});
	\item	Three frequency-based features (\texttt{freq\_05pc}, \texttt{freq\_med}, \texttt{freq\_95pc});
	\item	The ``Top-2'' performing features (\texttt{freq\_med}, \texttt{fm\_75pc});
	\item	All six FM and frequency-based features together.
\end{itemize}
We did not include the poorly-performing bandwidth features.
This yielded an advantage that the FM and frequency-based features had the same cardinality,
ensuring the fairness of our experimental comparison of the two feature types.

\begin{figure}[p]
	\centering
	\includegraphics [width=0.99\textwidth,clip,trim=1mm 0mm 10mm 17mm]  {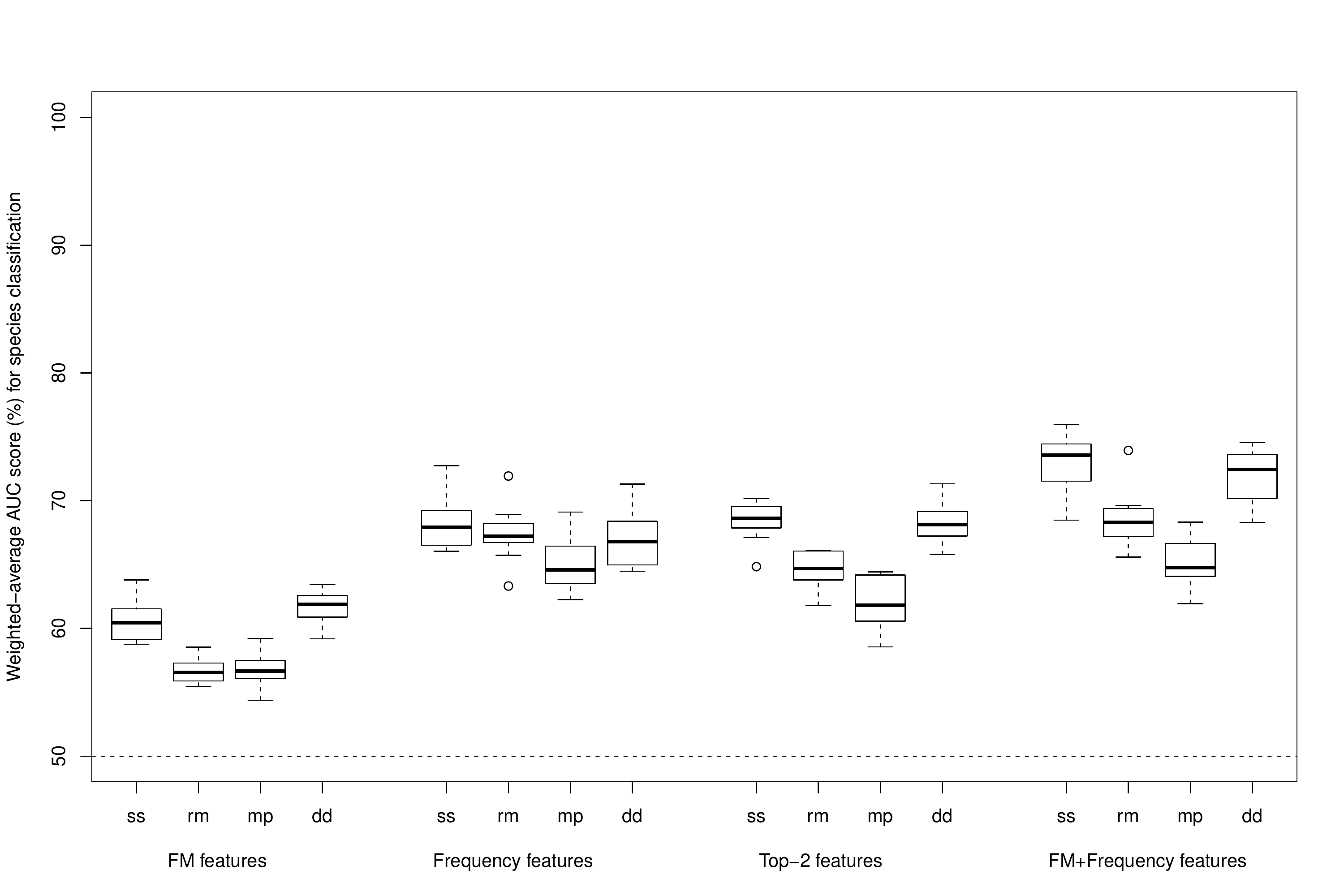}
	\includegraphics [width=0.99\textwidth,clip,trim=1mm 0mm 10mm 17mm]  {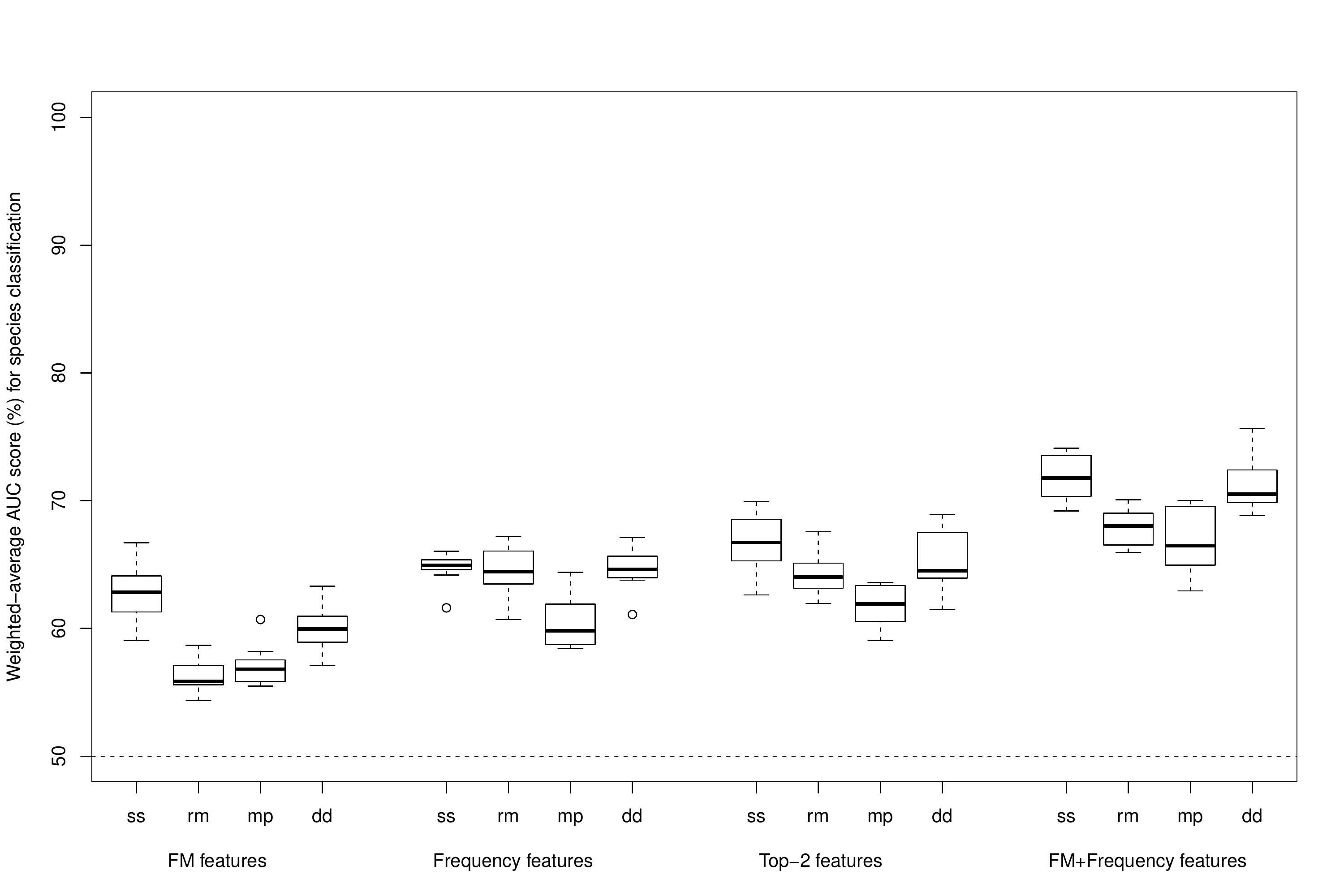}
	\caption{Performance of species classification across 56 species, evaluated using datasets \textit{PassaASA} (upper) and \textit{PhyllXC} (lower). %
	Results are shown for each analysis method, and for four different subsets of the available features (see text for details). %
	The horizontal dashed line indicates the baseline chance performance at 50\%. %
}
\label{fig:ssfbox}
\end{figure}

\begin{table}[t]
\caption{Marginal mean of the weighted area under the curve (AUC) scores for the results shown in Figure \ref{fig:ssfbox}.}
\label{tbl:ssfmeans}
	\centering
\begin{tabular}{l l l}
Dataset & Method & AUC (\%) \\
\hline
\textit{PassaASA} & \methss	&	\textbf{67.6}	\\
                  & \methdd	&	\textbf{67.2}	\\
                  & \methrm	&	64.3	\\
                  & \methmp	&	62.2	\\
\hline
\textit{PhyllXC} & \methss	&	\textbf{66.5}	\\
                  & \methdd	&	\textbf{65.3}	\\
                  & \methrm	&	63.2	\\
                  & \methmp	&	61.6	\\
\hline
& \\
Dataset & Feature set & AUC (\%) \\
\hline
\textit{PassaASA}  & FM+Freq	&	\textbf{69.6}	\\
                  & Top-2	&	65.8	\\
                  & Freq	&	66.9	\\
                  & FM  	&	58.9	\\
\hline
\textit{PhyllXC}  & FM+Freq	&	\textbf{69.5}	\\
                  & Top-2	&	64.4	\\
                  & Freq	&	63.6	\\
                  & FM  	&	59.1	\\
\hline
\end{tabular}
\end{table}

Results for the classification experiment with different extraction methods and different feature subsets
are shown in
Figure \ref{fig:ssfbox} and Table \ref{tbl:ssfmeans}.
This is a difficult classification task (across 56 species), and the average AUC score in this case peaks at around 70\%.
A repeated-measures factorial ANOVA confirmed, for both datasets, a significant effect on accuracy for both
feature set (\mbox{$p < 2 \times 10^{-16}$})
and
method (\mbox{$p \le 1.2 \times 10^{-6}$}),
with no significant interaction term found (\mbox{$p>0.07$}).

We conducted post-hoc tests for differences in AUC between pairs of methods and pairs of feature-sets,
using paired t-tests with Bonferroni correction for all pairwise comparisons
(this is a repeated-measures alternative to the Tukey HSD test).
Means were found to be different ($p < 0.0035$) for all pairs of methods except \methss vs.\ \methdd \mbox{(\methss $\approx$ \methdd $>$ \methrm $>$ \methmp)}.
For the choice of feature set,
means were found to be different ($p < 2.2 \times 10^{-6}$) for all pairs of feature sets except Top-2 vs.\ Freq \mbox{(FM+Freq $>$ Freq $\approx$ Top-2 $>$ FM)}.

\section{Discussion}
\label{sec:discussion}

The fine detail of frequency modulation (FM) is known to be used by various songbird species to carry information
(\citet[Chapter 7]{Marler:2004}; \citet{Brumm:2009,Sprau:2010,Vehrencamp:2013}),
but automatic tools for analysis of such FM are not yet commonly used.
Our experiments have demonstrated that FM information can be extracted efficiently from large datasets,
in a fashion which captures species-related information despite the simplicity of method
(we used no source-separation, syllable segmentation or pitch tracking).
This was explicitly designed for application on large collections:
our experiments used up to 1390 individual recordings, larger numbers than in many bioacoustic studies.


Our results show an effect of the choice of summary features, both for frequency and for FM data.
The consistently strongest-performing summary feature was the median frequency,
which is similar to measures of central tendency used elsewhere in the literature
and can be held to represent a bird's central ``typical'' frequency.
On the contrary, we were surprised to find that bandwidth measurements
as implemented in our study showed rather little predictive power for species identity,
since bandwidth has often been discussed with respect to the variation in vocal capacities across avian species \citep{Podos:1997,Trillo:2005,Mahler:2009}.
In our case the upper frequency extent alone (represented by the 95-percentile)
appears more reliable,
which may reflect the importance of production limits in the highest frequencies in song.

The FM features, taken alone, were not as predictive of species identity as were the frequency features.
However, they provided a significant boost in predictive power when appended to the frequency features.
This tells us not only that FM features encode aspects of species identity,
but they encode complementary information which is not captured in the frequency measurements.

In light of our results we note that \citet{Trillo:2005} explored a measure of ``trill vigour'':
``Because of the known production constraint trade-off between note rate and bandwidth of trilled songs (Podos 1997), we derived an index of trill vigour by multiplying the standardized scores of these two parameters'' \citep[p. 925]{Trillo:2005}.
This index was not further pursued since in their study it yielded similar results as the raw bandwidth data.
However, if we assume for the moment that each note in the trills studied by \citeauthor{Trillo:2005} is one full sweep of the bandwidth of the trill
(this is the case for all except ``hooked'' trills),
then multiplying the bandwidth (in Hz) by the note rate (in sec$^{-1}$) yields exactly the mean value of the instantaneous absolute FM rate (in Hz/sec).
This ``trill vigour'' calculation is thus very close in spirit to our measurement of the median FM rate.
Their comparison of bandwidth features against trill vigour features served for them as a kind of feature selection,
although in their case the focus was on trills in a single species.


A further aspect of our study is the comparison of four different methods for extracting FM data.
A clear result emerges from this, which is that the simplest method (\methssO)
attains the strongest classification results (tied with method \methddO),
and is sufficiently robust to the degradations we tested.
This should be taken together with the observation that it runs at least 20 times faster than any of the other methods on the same audio data,
to yield a strong recommendation for the \methss method.

This outcome came as a surprise to us,
especially considering the simplifying assumptions implicit in the \methss method.
It considers the peak-amplitude frequencies found in adjacent STFT frames (i.e.\ in adjacent ``slices'' of a spectrogram),
which may in many cases relate to the fundamental frequency of the bird vocalisation,
but can often happen to relate to a harmonic, or a chance fluctuation in background noise.
It contains no larger-scale corrections for continuity, as might be used in pitch-tracking-type methods
(though note that as we found with the method of \citet{Badeau:2006}, those methods can incur difficulties tracking fast modulations).

The statistical strength of simple methods has been studied elsewhere in the literature.
For example \citet{Kershenbaum:2013} found that bottlenose dolphin signature whistles
could usefully be summarised by a strongly decimated representation of the pitch track:
a so-called ``Parsons code'' based on whether the pitch is rising or falling at a particular timescale,
and which completely omits the magnitude of such rises or falls.
The method is not analogous to ours, but has in common that it uses suprisingly simple statistics to summarise temporal variation.
Audio ``fingerprinting'' systems such as Shazam \citep{Wang:2003} also rely on highly-reduced summary data,
customised to the audio domain of interest.

Our \methss method relies on finding a temporal difference between adjacent frames, as does that of \citet{Kershenbaum:2013}.
This is partly reminiscent of the ``delta'' features often added to MFCCs to reflect how they may be changing.
Such deltas are common in speech recognition and are also used in some automatic species classification
(for example \citet{Trifa:2008}).
However note that MFCC ``deltas'' represent differences in magnitude, not in frequency.


Separately from the classification experiment, we studied the effects of noise and MP3 degradation on our summary features.
Such issues are pertinent for crowdsourced datasets such as \textit{PhyllXC}.

Measures such as minimum and maximum frequency carry some risk of dependence on recording conditions,
particularly when derived from manual inspection of spectrograms \citep{Zollinger:2012,Cardoso:2012}.
We have demonstrated that our automatic FM measures using methods \methrmO, \methdd or \methss are robust against two common types of degradation (noise and compression), with \methrm particularly robust.
They are therefore suitable tools to explore the variation in songbirds' use of FM in the laboratory and in the field.

Future work: in this study we did not use any higher-level temporal modelling such as the temporal structure of trill syllables,
nor did we use advanced methods for segmenting song/call syllables from background.
We have demonstrated the utility of fully automatic extraction of fine temporal structure information,
and in future work we aim to combine this with richer modelling of other aspects of vocalisation.
We also look forward to combining fine FM analysis with physiological models of the songbird vocal production mechanism%
---as has already been done with linear prediction for the source-filter model 
\citep{Markel:1972}%
---but explicitly accounting for songbirds' capacity for rapid nonstationary modulation
and their use of two separate sound sources in the syrinx.

%% file: extraresults.tex
\clearpage
\section*{Appendix: Feature selection results}
\label{sec:extraresults}

\begin{figure}[h]
	\centering
	\includegraphics [width=0.99\textwidth,clip,trim=2mm 0mm 10mm 10mm]  {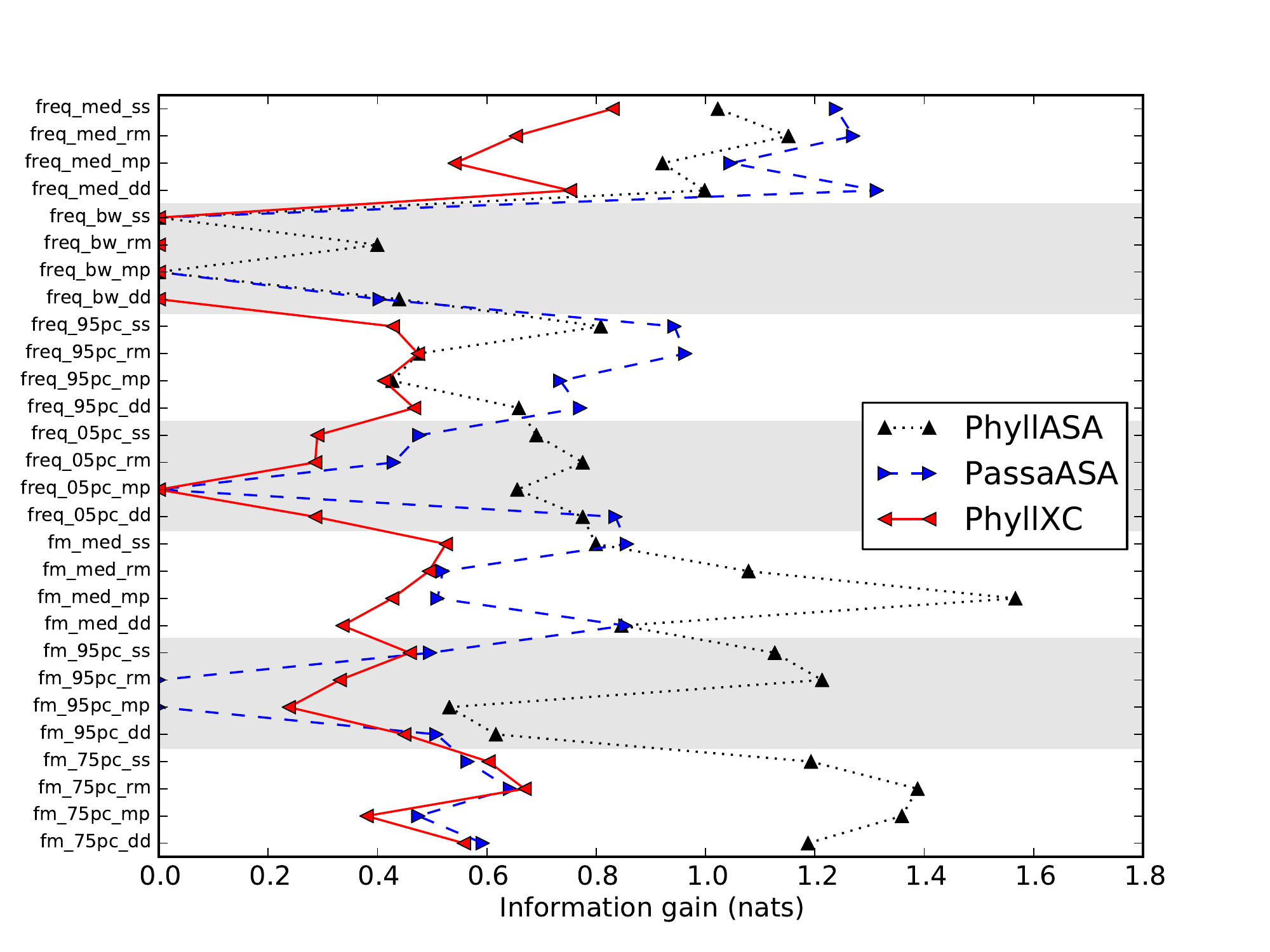}
	\caption{Overview of information gain (IG) values calculated during feature selection; %
	as in Table \ref{tbl:featsel} but ordered by feature type.%
	See Table \ref{tbl:featsel} for numerical values%
}
\label{fig:featselplot}
\end{figure}

\noindent We performed feature selection on each of our three datasets, evaluated using Information Gain (IG) as described in the main text
(Table \ref{tbl:featsel}, Figure \ref{fig:featselplot}).
The overall pattern of IG values shows broad similarities between \textit{PhyllASA} and \textit{PhyllXC},
indicating commonalities in species-dependent features.
The IG values evaluated on \textit{PhyllXC} are consistently lower than those in \textit{PhyllASA},
suggesting that the species information in the former may be affected by noise and MP3 effects.
However, the tendency to lower IG values may also be an artefact of differences in species distribution within each dataset.

 \begin{table}[h]
\caption{Ranked results of information-gain (IG) feature selection applied to each of our three datasets. %
Features are ranked in order of how strongly they predict species identity. %
Left to right: \textit{PhyllASA}, \textit{PassaASA}, \textit{PhyllXC}. %
}
\label{tbl:featsel}
\resizebox{!}{0.35\textwidth}{
	\begin{tabular}{l l l}
	Rank & IG & Feature \\
	\hline

	1	&	1.5667	&	fm\_med\_mp	\\
	2	&	1.3878	&	fm\_75pc\_rm	\\
	3	&	1.3591	&	fm\_75pc\_mp	\\
	4	&	1.2131	&	fm\_95pc\_rm	\\
	5	&	1.1928	&	fm\_75pc\_ss	\\
	6	&	1.1874	&	fm\_75pc\_dd	\\
	7	&	1.1516	&	freq\_med\_rm	\\
	8	&	1.1266	&	fm\_95pc\_ss	\\
	9	&	1.0786	&	fm\_med\_rm	\\
	10	&	1.0224	&	freq\_med\_ss	\\
	11	&	0.9984	&	freq\_med\_dd	\\
	12	&	0.9213	&	freq\_med\_mp	\\
	13	&	0.8461	&	fm\_med\_dd	\\
	14	&	0.8084	&	freq\_95pc\_ss	\\
	15	&	0.7994	&	fm\_med\_ss	\\
	16	&	0.7754	&	freq\_05pc\_rm	\\
	17	&	0.7754	&	freq\_05pc\_dd	\\
	18	&	0.6906	&	freq\_05pc\_ss	\\
	19	&	0.6587	&	freq\_95pc\_dd	\\
	20	&	0.6556	&	freq\_05pc\_mp	\\
	21	&	0.6165	&	fm\_95pc\_dd	\\
	22	&	0.5314	&	fm\_95pc\_mp	\\
	23	&	0.4748	&	freq\_95pc\_rm	\\
	24	&	0.4396	&	freq\_bw\_dd	\\
	25	&	0.4273	&	freq\_95pc\_mp	\\
	26	&	0.3998	&	freq\_bw\_rm	\\
	27	&	0	&	freq\_bw\_mp	\\
	28	&	0	&	freq\_bw\_ss	\\
	\hline

	\end{tabular}
}
\resizebox{!}{0.35\textwidth}{
	\begin{tabular}{l l l}
	Rank & IG & Feature \\
	\hline
	1	&	1.3133	&	freq\_med\_dd	\\
	2	&	1.2701	&	freq\_med\_rm	\\
	3	&	1.2387	&	freq\_med\_ss	\\
	4	&	1.0457	&	freq\_med\_mp	\\
	5	&	0.9629	&	freq\_95pc\_rm	\\
	6	&	0.9432	&	freq\_95pc\_ss	\\
	7	&	0.8563	&	fm\_med\_ss	\\
	8	&	0.8533	&	fm\_med\_dd	\\
	9	&	0.8353	&	freq\_05pc\_dd	\\
	10	&	0.7708	&	freq\_95pc\_dd	\\
	11	&	0.7343	&	freq\_95pc\_mp	\\
	12	&	0.6424	&	fm\_75pc\_rm	\\
	13	&	0.5923	&	fm\_75pc\_dd	\\
	14	&	0.5648	&	fm\_75pc\_ss	\\
	15	&	0.5194	&	fm\_med\_rm	\\
	16	&	0.5098	&	fm\_med\_mp	\\
	17	&	0.5079	&	fm\_95pc\_dd	\\
	18	&	0.4964	&	fm\_95pc\_ss	\\
	19	&	0.4767	&	freq\_05pc\_ss	\\
	20	&	0.4747	&	fm\_75pc\_mp	\\
	21	&	0.43	&	freq\_05pc\_rm	\\
	22	&	0.4039	&	freq\_bw\_dd	\\
	23	&	0	&	freq\_bw\_rm	\\
	24	&	0	&	fm\_95pc\_rm	\\
	25	&	0	&	freq\_bw\_mp	\\
	26	&	0	&	fm\_95pc\_mp	\\
	27	&	0	&	freq\_bw\_ss	\\
	28	&	0	&	freq\_05pc\_mp	\\
	\hline
	\end{tabular}
}
\resizebox{!}{0.35\textwidth}{
	\begin{tabular}{l l l}
	Rank & IG & Feature \\
	\hline
	1	&	0.83	&	freq\_med\_ss	\\
	2	&	0.752	&	freq\_med\_dd	\\
	3	&	0.669	&	fm\_75pc\_rm	\\
	4	&	0.653	&	freq\_med\_rm	\\
	5	&	0.603	&	fm\_75pc\_ss	\\
	6	&	0.558	&	fm\_75pc\_dd	\\
	7	&	0.541	&	freq\_med\_mp	\\
	8	&	0.525	&	fm\_med\_ss	\\
	9	&	0.494	&	fm\_med\_rm	\\
	10	&	0.474	&	freq\_95pc\_rm	\\
	11	&	0.467	&	freq\_95pc\_dd	\\
	12	&	0.459	&	fm\_95pc\_ss	\\
	13	&	0.449	&	fm\_95pc\_dd	\\
	14	&	0.428	&	freq\_95pc\_ss	\\
	15	&	0.427	&	fm\_med\_mp	\\
	16	&	0.412	&	freq\_95pc\_mp	\\
	17	&	0.38	&	fm\_75pc\_mp	\\
	18	&	0.336	&	fm\_med\_dd	\\
	19	&	0.331	&	fm\_95pc\_rm	\\
	20	&	0.29	&	freq\_05pc\_ss	\\
	21	&	0.286	&	freq\_05pc\_rm	\\
	22	&	0.286	&	freq\_05pc\_dd	\\
	23	&	0.238	&	fm\_95pc\_mp	\\
	24	&	0	&	freq\_bw\_dd	\\
	25	&	0	&	freq\_bw\_mp	\\
	26	&	0	&	freq\_05pc\_mp	\\
	27	&	0	&	freq\_bw\_ss	\\
	28	&	0	&	freq\_bw\_rm	\\
	\hline
	\end{tabular}
}
\end{table}

